\documentclass[12pt, review]{article}
\pdfoutput=1
\usepackage{lineno,hyperref,todonotes}
\usepackage{xcolor}
\usepackage{soul}
\usepackage{authblk}

\newcommand{\DipInfHack}{This work was supported in part by the MIUR under grant "Dipartimenti di eccellenza 2018-2022" of the Department of Computer Science of Sapienza University.}

\modulolinenumbers[5]
\begin{document}


\title{ \textbf{Quality of Life Assessment of Diabetic patients from health-related blogs} }

\author[1]{Andrea Lenzi}
\author[2]{Marianna Maranghi}
\author[3]{Giovanni Stilo}
\author[1]{Paola Velardi \thanks{Corresponding author: velardi@di.uniroma1.it}}

\affil[1]{ \textit{\footnotesize Sapienza University of Rome - Computer Science Departement} }
\affil[2]{ \textit{\footnotesize Sapienza University of Rome - Department of Translational and Precision Medicine} }
\affil[3]{ \textit{\footnotesize University of L'Aquila - Department of Engineering and Information Science and Mathematics} }  

\date{{\small \today}}

\maketitle

\begin{abstract}
 \textit{Motivations:} People are  generating an enormous amount of social data to describe their health care experiences, and continuously search  information about diseases, symptoms, diagnoses, doctors, treatment options and medicines. The increasing availability of these social traces presents an interesting opportunity to enhance timeliness and efficiency of care. By collecting, analyzing and exploiting this information, it is possible to modify or in any case significantly improve our knowledge on the manifestation of a pathology and obtain a more detailed and nuanced vision of  patients' experience, that we call the  "social phenotype" of diseases.\\
\textit{Materials and methods:} In this paper we present a data analytic framework to  represent, extract and analyze the social phenotype of diseases.
To show the effectiveness of our methodology we presents a detailed case study on diabetes. First, we create a high quality data sample of  diabetic patients' messages, extracted from popular medical forums during more than 10 years. Next, we use a topic extraction techniques based on latent analysis and word embeddings, to identify the main complications, the frequently reported symptoms and the common concerns of these patients.\\
\textit{Results:} We show that a freely manifested perception of a disease can be  noticeably different from what is inferred from  questionnaires, surveys and other common methodologies used to measure the impact of a disease on the patients' quality of life. In our case study on diabetes, we found that  issues reported to have a daily impact on diabetic patients are diet, glycemic control, drugs and clinical tests. These problems  are not commonly considered in Quality of Life assessments, since they are not perceived by doctors as representing severe limitations.  
\end{abstract}

\providecommand{\keywords}[1]{\textbf{\textit{Keywords:}} #1}



\section{Introduction}
\label{intro}
On-line patient health data  is increasing exponentially. Every day, through social networks, search engines, online forums, web sites, mobile applications, IoT devices, etc., people generate an enormous amount of health information \cite{ref:4}. This digital data (often referred to as the "digital phenotype" \cite{2015fOellrich}) includes symptoms, medical experiences, bio data, biometric information, web search queries, personal stories and medical records. They represent the footprints of the patient’s health status and 
foster large-scale and low-cost analysis of  common diseases \cite{ref:2} for many applications, such as syndromic surveillance \cite{Velardi:2014:TMF:2657512.2657841}, adverse drug reactions \cite{DBLP:conf/amia/OConnorNGPSSG14}, and "social phenotyping" of diseases, i.e., a \textit{subjective and unfiltered picture of  perceptions, health conditions, feelings and lifestyles} of people affected by a given disease\footnote{Note that our notion of social phenotype is more focused than that of digital phenotype. The latter includes a variety of digital data, the former is  concerned only with social traces.}. 
The social phenotye has a value that goes beyond the  physical, laboratory and clinical imaging examinations, which represent the traditional approaches to define the phenotype of a disease. By collecting, analyzing and integrating this information in an appropriate manner, it is possible to obtain a more detailed and "nuanced" vision of patients,  taking into account different expressions of diseases; the experience lived outside the traditional syndromic care and monitoring structures;  the  symptoms that are perceived as more limiting; the greater or lesser tolerance to drugs. This information is also very useful to complement  standard survey-based techniques to assess patients' Quality of Life (QoL). Although QoL surveys are very useful in assessing patients' evaluations of care delivery,   administered questionnaires reflect the care-givers' view of the factors that mostly influence  the health status and everyday life, while the social phenotype fosters a patient-centric perspective. 

Leaning on these concepts, in this paper we gathered an extensive amount of health-related textual data from four popular medical forums. In these web  platforms, patients submit health problems, questions and personal  experiences to doctors. We applied text analytics and topic detection on these data with the purpose of answering the following research questions:
\begin{enumerate}
    \item R1: What are the main complications reported by patients for selected diseases?  Which symptoms  are more frequently reported by patients? 
    \item R2: Are there gender/age differences in a disease perception?
    \item R3: Does this "patient-centric"  study of a disease perception correspond to published results using traditional Quality of Life assessments?
\end{enumerate}

The paper is organized as follows: In Section \ref{sec:related} we summarize related literature. In Section \ref{sec:SPEP} we describe the process of data collection and filtering, and the adopted topic extraction algorithm. Section \ref{sec:phenotype}  is dedicated to the analysis of the social phenotype of a specific and highly diffused disease, diabetes. Finally, Section \ref{sec:conclusion} presents a summary of findings and concluding remarks.

\section{Related work}
\label{sec:related}

Recent years have seen increasing interest in patient-centered care and in the improving of the patient experience. At the same time, through the Internet, people are generating an enormous amount of data to describe their experiences of healthcare \cite{ref:2}. In fact, more and more often patients use the Web 2.0 technologies to search for information about diseases, symptoms, diagnoses, doctors, treatment options and medicines. They also search for support from online communities and post states about their conditions and their sensations \cite{ref:10}. The increasing availability of this information presents an interesting opportunity to enhance timeliness and efficiency of care. Greaves et al. \cite{ref:2} describe this growing body of information on the Internet as a ‘cloud of patient experience’. 

Jain et al. \cite{ref:3} starting from Richard Dawkins' idea of an extended phenotype (“phenotypes should not be limited just to biological processes, but extended to include all effects that a gene has on its environment inside or outside of the body of the individual organism” \cite{ref:3}) introduced the concept of "digital phenotype". Through technology, including social media, search engines, online communities, forums, blogs, wearable technologies and mobile devices, people are generating an enormous amount of health-related data that can provide deep knowledge of patient’s health. These sources of information give us the footprints of the patient’s health status and could enrich notion of extended phenotype in a substantial way. Precisely, when gathered and analyzed suitably, such data provide a more comprehensive and nuanced view of the experience of disease (from diagnosis to treatment) and expand our ability to identify and diagnose health conditions.

Leaning on these concepts, in the last years, several studies showed that it was possible through machine learning techniques to detect patterns in data gathered from forums or social networks related to patient experience, in order to make decisions or predictions. Furthermore, there is an ample literature (see \cite{surveyblogs} for a survey) that demonstrates the possibility to analyze different types of discussion forums or blogs for obtaining useful information. Recent work more related to the objectives of this paper  is summarized in what follows.

Verberne et al. \cite{ref:11} investigate the feasibility of thread summarization in patients' support forum, selecting the most relevant sentences
from a thread while hiding the less relevant sentences. They collected two sets of reference summaries for 100 randomly selected threads, with at least 10 posts, from the Facebook group of a patient support community: expert summaries and crowdsourced summaries. For each discussion thread they calculated the agreement between each pair of raters in terms of Cohen’s k. They found that the inter-rater agreement in crowdsourced summaries is low, close to random (k=0.081); while the inter-rater agreement in expert summaries achieved a considerably higher, fair value (k=0.267). Afterward, they selected the number of votes for a sentence as dependent variable and features related to frequency, similarity and position of the words as independent variables. They trained two linear regression models using the two types of reference summaries and the selected variables. The models looked to be similar. So they concluded that it is possible to train a summarization model on crowdsourced information that is similar to an expert model, even if the inter-rater agreement for the crowdsourced summaries is very low.

White at al. \cite{ref:12} examined the use of web search behaviors for detecting neurodegenerative disorders (ND), such as Parkinson’s disease (PD) and Alzheimer’s disease (AD) from search activities. They used a total of 18 months (from September 2015 to February 2017) of anonymous logs of United States search activity from the Microsoft Bing web search engine, containing millions of English-speaking searchers (31\,321\,070). From the full set of logs, searchers who input queries containing first-person statements about PD diagnosis were used as evidence of receiving a PD diagnosis. Multiple additive regression trees (MART) classifiers were trained to detect evidence of PD diagnosis. 10-fold cross-validation was used to train and test the classifier. Classifier sensitivities for PD discovery were 94.2/83.1/42.0/34.6\% at false positive rates (FPRs) of 20/10/1/0.1\% respectively (AUROC=0.9357). Furthermore, the method in this study was scaled to Alzheimer’s disease. In this case, sensitivities for AD detection were 91.0/81.5/38.8/26.1\% at false positive rates (FPRs) of 20/10/1/0.1\% respectively (AUROC=0.9135). The results offered evidence that the existence of NDs in web searchers was detectable from streams of data through an analysis over time.

Mirheidari et al. \cite{ref:13} using conversation analysis (CA), showed that a machine learning approach to analyze transcripts of interactions between neurologists and patients, who described memory problems, can differentiate people with neurodegenerative memory disorders (ND) from people with functional memory disorders (FMD). Firstly, 15 patients with ND and 15 patients with FMD, whose diagnoses could be established with wide certainty by the doctors, were selected for the study. Later, an automatic speech recognition (ASR) system and a diarization tool produced the written records of conversations. The output of the ASR+Diarization was passed on to the CA-style feature extraction unit which selected features related to the time of a conversation's turn, the pauses in speech and the words used. Afterward, the features extracted were used to train a set of five binary classifiers to differentiate between ND and FMD. Each classifier reached a significant level of accuracy (Perceptron: 97\%; Linear SVM: 93\%; AdaBoost: 93\%; Linear via SGD: 93\%; Random Forest: 90\%).


Miller et al. \cite{ref:10} analyzed the content and characteristics of important health blogs to supply a more complete understanding of the health blogosphere. They identified, through a purposive-snowball sampling approach, 951 health blogs during two periods: June to July 2007 and April to May 2008. All blogs were US focused, written in English and updated frequently. They selected as features for the analysis: main blog topics, demographic information, frequency of posts, presence or absence of links and comment section. Most blogs typically focused on bloggers’ experiences. Half of the blogs were written from a professional view, one-third from a patient perspective. The majority of the bloggers were female, aged in their 30s and highly educated. So, in conclusion, they proved that data gathered from health blogs could be aggregated for large-scale experimental investigations.

Kavakiotis et al. \cite{ref:1} starting from the fact that diabetes mellitus is rapidly emerging as one of the greatest international health challenges of the 21st century, conducted a systematic survey of applications of data analytic techniques on scientific papers regarding diabetes. They identified and reviewed data mining and machine learning approaches applied on diabetes research respect to these five themes: “Prediction and Diagnosis through Biomarker Identification”, “Diabetic Complications”, “Drugs and Therapies”, “Genetic Background and Environment” and “Health Care and Management Systems”. In general, 85\% of those researches were conducted by supervised learning approaches and 15\% by unsupervised techniques; Support Vector Machines (SVM) represented the most successful and widely used algorithm; clinical data-sets were principally used; the primary purpose of these studies was to extract valuable knowledge that led to new hypotheses aimed at a deeper understanding and further investigation into diabetes.

The literature review on patients' blogs analytics shows that the majority of works concentrated on classification problems rather than  deriving a comprehensive picture of patient's perceptions and lifestyle. Although the idea of extracting the digital phenotype of diseases has been launched in \cite{ref:3}, to the best of our knowledge, this is the first large-scale study in which the utility of this approach is demonstrated.
\section{The Social Phenotype Extraction Process}
\label{sec:SPEP}
\begin{figure}
  \centering
  \includegraphics[width=0.9\linewidth]{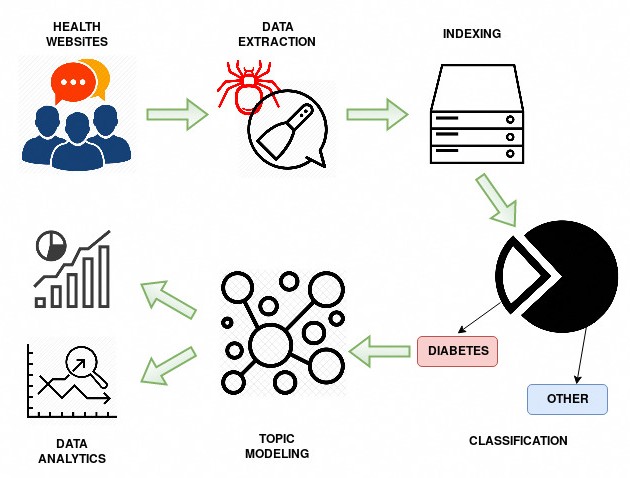}
  \caption{\textit{\small Workflow of the Social Phenotype Extraction Process}}
  \label{fig:generalScheme}
\end{figure}

This Section describes in more detail the processing steps  to extract the Social Phenotype of a disease, shown in the workflow of Figure \ref{fig:generalScheme}:  
\begin{enumerate}
    \item First, we collected a large number of threads from popular medical blogs in Italy from 2001 to 2018 (Section (\ref{sec:data}); 
    \item Secondly, we  filtered from these blogs conversations related to diabetes (Section \ref{subsec:filtering}). Diabetes is one of the most common endocrine disorders, affecting hundreds of million of people globally; furthermore, the number of diabetic people is estimated to grow dramatically in the upcoming years \cite{ref:4}, also due to the increase in obesity and sedentarity.  
\item Next, we applied a topic extraction technique based on latent analysis augmented with word embeddings (Section  \ref{subsec:topic}). Automatically extracted topics were found to nicely identify the main known complications of diabetes, complained symptoms, effects of drugs, and clinical tests, although no prior knowledge was provided to the topic learner.
For a subset of messages, we have been able to infer personal data such as gender and age, using specific keywords (Section \ref{subsec:genderage});
\item Finally, data analytics has been extensively carried on, to identify co-morbidities, symptoms, conditions and main patients' concerns along different dimensions, such as type of complication, age and gender, thus deriving a synthetic representation of diabetic patients' experience in their everyday life, which we call the social phenotype of a disease (Section \ref{sec:phenotype}). 
\end{enumerate}

\subsection{Dataset creation}
\label{sec:data}
We scraped a large set of data (about 700 thousand threads containing more than 3 million messages) from the 4 most popular medical forums in Italy, \textit{medicitalia.it}, \textit{forumsalute.it}, \textit{paginemediche.it} and \textit{pazienti.it} \footnote{Patients are anonymous and we have followed the current privacy regulations according also with the T.O.S. of the involved platforms.}.  Conversations span from 2001 to June 2018, although the large majority of messages refers to the past 10 years. The total number of threads and messages is summarized in Table \ref{dataset:statistics}. 


\begin{table}[tbp]
\centering            
\caption{\textit{\small Total number of threads and messages}}
\label{dataset:statistics}
\begin{tabular}{|l|l|l|}
\hline
 \textbf{source} & \textbf{number of threads} & \textbf{number of messages} \\ \hline \hline
\textbf{\textit{medicitalia.it}} & 562\,817 & 2\,597\,881  \\ \hline
\textbf{\textit{forumsalute.it}} & 38\,404 & 431\,480 \\ \hline
\textbf{\textit{paginemediche.it}} & 81\,373 & 162\,746  \\ \hline
\textbf{\textit{pazienti.it}} & 14\,853 & 29\,706  \\ \hline
\hline
\textbf{Total} & 697\,447 & 3\,221\,813  \\ \hline
\end{tabular}
\end{table}

In this work, we decided to focus on a specific pathology: diabetes.  We have chosen to examine  the social phenotype of this widespread and problematic metabolic disease considering the following alarming statistics and evaluations:

\begin{itemize}
\item diabetes interests more than 415 million people worldwide;
\item 29.1 million of U.S. citizens had diabetes in 2012 (9.3\% of the population);
\item the proportion of the people with diagnosed diabetes is expected to increase (1 in 10 adults will have diabetes by 2040, and 1 in 3 adults of USA will have diabetes by 2050);
\item diabetes care has been estimated at \$1.31 trillion globally;
\item diabetic patients have a 3-fold greater possibility of hospitalization in comparison to patients without this illness;
\item for the same condition, people hospitalized with a diagnosis of diabetes stay in the hospital for more time than patients without a diagnosis of this disease;
\item this pathology is the 7th cause of death in the USA in 2014 (24 deaths each 100\,000 people);
\item hyperglycemia is strongly connected with an increased risk of mortality and complications \cite{ref:3}.
\end{itemize}
In Italy, the picture is very similar, if not worst:
the prevalence of diabetes nearly doubled in thirty years.  Compared to 2000, there are 1 million more people with diabetes and this is due both to the aging of the population and to other factors, including the anticipation of diagnoses (which brings to light previously unknown cases) and the increase in survival of patients with diabetes.

\subsection{Classification of diabetic patients}
\label{subsec:filtering}
Although most threads have category labels indicating the discussion topic, and although diabetes is one of the categories, after a closer inspection by a diabetologist (one of the co-authors) we concluded that the classification was often not accurate. First, some of the threads manually tagged  as "diabetes" actually show a different symptomatology (although the patient fears to have diabetes symptoms), second, and more importantly, several diabetes-related threads are classified with other category labels, reflecting common diabetes complications and risk factors, such as cardiology, obesity, and others. For this reason it has been necessary to re-classify all threads using automated, but eventually more performant, techniques.  

To classify messages, we used the open-source library \textit{FastText}. This software tool, developed by \textit{Facebook's Artificial Intelligence Research (FAIR)} lab, combines state-of-the-art  Natural Language Processing  and Machine Learning techniques
\cite{ref:15}. The classification algorithm consists in a fast approximation of the \textit{softmax classifier}, in which it is used a hierarchical structure (a tree, representing categories, assembled by using the Huffman algorithm) instead of a flat construction \cite{ref:15}. 

In this phase of our study, we devoted specific attention to the training and validation process, \textit{since we considered of particular importance to obtain a clean dataset of conversations clearly associated to diabetes,  its complications and risk factors}. 

The \textit{training set} was built as follows: we selected the threads that contained at least three times the prefixes "diab" and "insulin" (the roots of the most characterizing words of this specific disease) and we assigned to them the label "DIABETES". We adopted this approach, since it was manually verified on a sub-sample that, in this way,   only the threads strongly connected to the pathology of interest were chosen. Next, we retrieved a dictionary of terms (key-words and key-phrases) related to diabetes from the site \href{http://www.diabete.net/conoscere-il-diabete/dizionario/}{\textit{diabetes.net}}, we selected threads that did not contain any of these words homogeneously between categories, and we confer to these documents the label "NO".  
At the end of these operations, the resulting training set consisted of 380\,000 threads, of which: 
\begin{itemize}
\item 3\,902 threads labeled as "DIABETES";
\item 376\,098 threads labeled as "NO".
\end{itemize}
Once the training set was built, we  trained the model using  various combinations of hyper-parameters. Subsequently, we selected the model that allowed us to achieve the best performance in the automatic evaluation tests. 
Finally, we used the best learned model to classify  all the documents in the initial collection. \textit{At the end, 10\,388 threads, containing a total of 40\,443 patients'  messages\footnote{Note that threads often include conversations between patients and doctors. Doctors' messages, however, are clearly marked as such. For the purpose of this analysis, doctors' messages have been deleted.}, were classified as related to the pathology of interest}. 

To ensure data quality, we applied multiple validation techniques on this dataset. 
We first evaluated performances using the classic holdout method (70\% training, 30\% testing).  With this method we detected no errors (zero False-Positives and zero False-Negatives).
Next, we repeated automated evaluation using a \textit{validation set}  of discussion threads extracted from another specific forum  exclusively dealing with diabetes (\href{http://www.portalediabete.org/esperti/faq}{\textit{portalediabete.org}}) and we marked the 506 gathered threads  with the label “DIABETES”. Subsequently, we extracted discussion threads from other forums related to medical areas not connected to diabetes on the site \href{http://www.abcsalute.it/forum/}{\textit{abcsalute.it}}, selecting only  items that contained no terms strictly characterizing the disease. We assigned to the 2\,345 threads obtained in this way the label "NO". Finally, with this validation set of 2\,851 instances, we tested the previously obtained classification model. The   performances  are shown in Table \ref{performanceAutomatic}.

\begin{table}[tbp]
\centering
\caption{\textit{\small Classification performances on the validation set}}
\label{performanceAutomatic}
\begin{tabular}{|l|l|}
\hline
\textbf{metric} & \textbf{score} \\ \hline
Accuracy & 0.993 \\ \hline
Error rate & 0.007 \\ \hline
Precision & 1 \\ \hline
Recall (TPR) & 0.960 \\ \hline
Specificity (TNR) & 1 \\ \hline
F-measure & 0.980 \\ \hline
\end{tabular}
\end{table}

Lastly, we performed manual evaluation by experts.
We sampled 500 threads from the original test set, with the predictions made by the model. 250 items in the sample presented the label "DIABETES", while the remaining 250 threads were labeled "NO". We then asked two diabetologists to manually evaluate the correctness of these predictions. We exercised three different degrees of judgment (Best case, Average case and Worst case) based on the severity of reporting False-Positives errors. For example, in the \textit{Average case} we considered True-Positives all threads clearly connected to diabetes, or otherwise that presented in their messages the following elements: problems of glucose metabolism, metabolic study using a glycemic and insulinemic curve, metabolic syndrome, hyperglycemia, analysis of blood sugar and insulinemia, diabetes as a risk factor in the answer provided by the doctor, specific drugs to treat this disease.


For all three levels of judgment, the performances of the classification obtained from manual evaluation are listed in Table \ref{performanceManual}. In conclusion, according to the performed evaluations, it can be stated that the quality of the classification task resulted satisfactory and adequate for our context. The dataset of 40\,443 patients'  messages was therefore finally adopted.

\begin{table}[tbp]
\centering            
\caption{\textit{\small Classification performance obtained from manual evaluation}}
\label{performanceManual}
\begin{tabular}{|l|l|l|l|}
\hline
 & \textbf{Average case} & \textbf{Worst case} & \textbf{Best case} \\ \hline
\textbf{Accuracy} & 0.950 & 0.920 & 0.976 \\ \hline
\textbf{Error rate} & 0.050 & 0.080 & 0.024 \\ \hline
\textbf{Precision} & 0.900 & 0.840 & 0.952 \\ \hline
\textbf{Recall (TPR)} & 1 & 1 & 1 \\ \hline
\textbf{Specificity (TNR)} & 0.909 & 0.862 & 0.954 \\ \hline
\textbf{F-measure} & 0.947 & 0.913 & 0.975 \\ \hline
\end{tabular}
\end{table}

\subsection{Topic detection}
\label{subsec:topic}

Our objective in this research is  to characterize  patients' experience and focus on how they perceive the disease and its complications. To this end, topic extraction has been applied solely on patients' messages in our collection (doctors' messages have been excluded).  In this context, a \textit{topic} is a set of semantically and/or contextually related words or key-phrases, representing a coherent and frequent argument of discussion.

We adopted a hybrid approach to discover the main discussion topics of diabetic patients. First, we applied LDA \cite{ref:19} on the collection and then, we improved the quality of obtained clusters using the Word2Vec model.

Latent Dirichlet allocation (LDA)  is a three-level hierarchical Bayesian model, in which each document is represented as a finite mixture, according to a Dirichlet distribution, over an underlying fixed set of topics. In turn, each topic is modeled by a probability distribution over words \cite{ref:19}.
Word2Vec is a computationally efficient family of predictive models for learning word embeddings from large amounts of unstructured text data, introduced by Mikolov et al. \cite{ref:20}. Word2Vec consists in a multi-layer neural network. It takes as its input raw text and produces a continuous vector space, in which each single word in the corpus is mapped onto a corresponding vector of real numbers in a lower-dimension semantic space. Semantically similar vectors are located in close proximity to one another in the vector space. 
To compute both LDA and Word2Vec, we employed \textit{Gensim}, an open-source Python library for topic extraction and vector space modeling.

In the \textit{LDA phase},  we calculated the term-document matrix from the collection of documents regarding diabetes, subsequently we processed and normalized this data structure, and finally we trained LDA with this matrix as input. To normalize  the  values in the matrix, we used as weight function the mean between TF-IDF and log-entropy models. The \textit{Log-Entropy model} is a more sophisticated weighting scheme based on entropy. This model operates row-wise with no interaction between terms and produces normalized values in the range from zero to one. In this scheme, a reduced weight is assigned to a word that appears with equal probability between documents; while a more significant weight is given to a term whose occurrence is concentrated in few documents.
\begin{displaymath} 
local\_weight_{t,d} = \log{ \left( TF_{t,d} + 1 \right) }
\end{displaymath}

\begin{displaymath}
global\_weight_{t} = 1 + \frac{\sum_{d} \left(\frac{TF_{t,d}}{\sum_{d}TF_{t,d}} \cdot \log{ \left( \frac{TF_{t,d}}{\sum_{d}TF_{t,d}} \right) } \right )  } {\log \left( N +  1  \right)}
\end{displaymath}

\begin{displaymath}
final\_weight_{t,d} = local\_weight_{t,d} \cdot global\_weight_{t} 
\end{displaymath}


At the end of this phase, we obtained 14\footnote{the number of topics is an hyper-parameter of the LDA model. This number is commonly  tuned experimentally in the LDA literature.} "temporary" topics, which where then refined in the subsequent phase.

During the \textit{Word2Vec phase}, we trained the \textit{CBOW} model of the Word2Vec family with  the entire collection of discussion threads (concerning all types of diseases).  We obtained a continuous vector space, able to model the contextual similarities and differences between words in the collection.
We trained the Word2Vec model with the following parameters:
\begin{itemize}
\item \textit{min\_count} = 100 (minimum frequency: the algorithm neglects all words with number of occurrences lower than the specified value);
\item \textit{size} = 100 (Dimensionality of the word vectors);
\item \textit{iter} = 100 (Number of iterations over the corpus).
\end{itemize}

In the last phase, we  improved the quality of the 14  topics previously obtained with LDA, using word embeddings learned during phase 2. For this purpose, we initially removed 30\% of the least relevant words from each temporary cluster $T_i$ according to Word2Vec. Then, we added to each $T_i$ all the words that Word2Vec considered similar to those in $T_i$ with a similarity score higher than 0.60\footnote{these hyper-parameters have been experimentally tuned to obtain top quality topics, according to domain experts.}. At the end of these operations, we obtained 14  improved topics able to model quite accurately various aspects of the pathology under study.

\begin{figure}[h]
  \centering
  \includegraphics[width=1.05\linewidth]{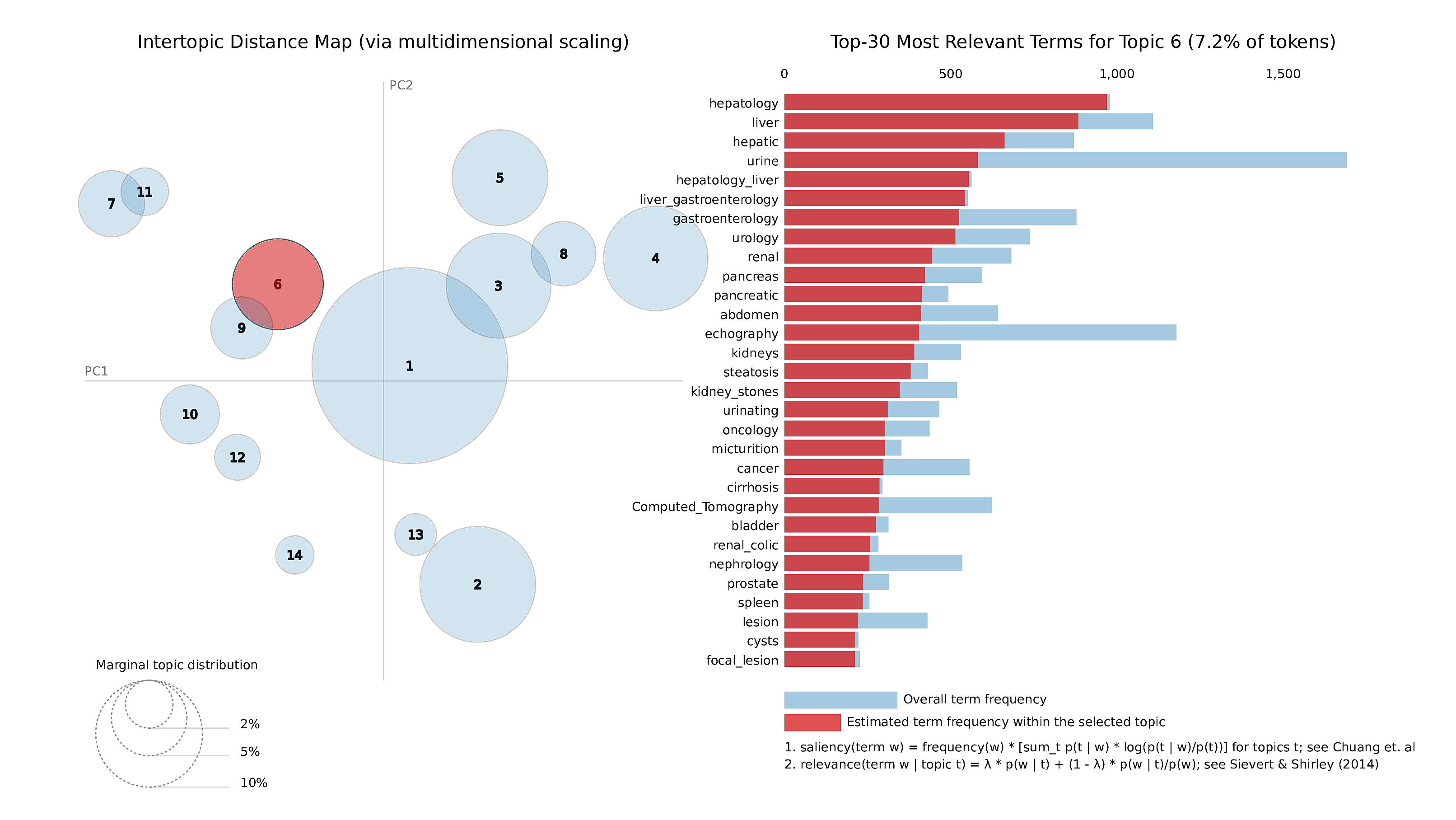}
  \caption{\textit{\small A visualization of topic 6, Hepathology}}
  \label{fig:topics6}
\end{figure}

Figure \ref{fig:topics6} shows graphically the relevance of each topic and their intersections. In this view, the topics are plotted as circles in a two-dimensional plane. Topic centers are discovered by calculating the distance between  topics, and then by using multidimensional scaling to project the inter-topic distances on the two dimensions. Moreover, each topic prevalence is encoded using the areas of the circles \cite{ref:22}. 

Figure \ref{fig:topics6} also shows a horizontal bar-chart representing the 30 most salient terms in the diabetes collection related to a selected topic  (in the Figure, the highlighted topic is topic 6, labeled \textit{hepathology}). Saliency is based on the following probabilities: 

\begin{itemize}
\item $ \Pr \left( w \right) $ represents the probability of randomly selecting the word \textit{w} from the collection (the frequency of the term \textit{w} normalized in the range between zero and one),
\item $ \Pr \left( T \mid w \right) $ represents the likelihood that observed word \textit{w} was generated by latent topic \textit{T},
\item $ \Pr \left( T \right) $ represents the likelihood that any randomly-selected word ${w}'$ was generated by topic \textit{T};
\end{itemize}

the saliency of a word \textit{w} is then defined by this equation \cite{ref:23}:

$$ saliency\left(w\right) = \Pr\left(w\right) \cdot \sum_T \left(  \Pr\left(T \mid w\right) \cdot \log \left(  \frac{\Pr\left(T \mid w \right)}{\Pr\left(T\right)} \right) \right) $$

The 14 extracted topics are listed in Appendix 1 in descending order from the most frequent to the least predominant in the collection. For all topics, only the most significant terms are reported, in order of importance. All terms have been translated from Italian to English for readability. After a meticulous manual evaluation\footnote{Note however that we did \textit{not} manually adjusted the list of words in each topic: although there might be a few apparently spurious words, they still reflect similarity of usage, according to the adopted model.}, we assigned to each topic the label that seemed to be more appropriate and representative. 
 
Appendix 1 shows that 9 clusters nicely\footnote{Remember that topic clustering is an untrained methodology: no  domain knowledge is available to the system} correspond to the main known complications of diabetes (clusters 2, 4, 6, 7, 9, 10, 11, 12, 13 and 14); cluster 1 includes  terms related to feelings and symptoms; cluster 3 is about dietary issues; cluster 5 is about clinical tests and drugs; cluster 8 includes  terms related to diabetes in general (e.g. mellitus, insulin, glycemia, etc.) and finally, cluster 9 has no clear semantics\footnote{"Everything else" clusters are an almost unavoidable phenomenon in clustering, although in our case it is a small cluster.}. As shown in Figure \ref{fig:topics6}, the largest cluster is cluster 1, labeled \textit{symptoms-feelings}. This is the most important cluster, both because of its frequency, and because of the objectives of our analysis. For this reason, it has been further partitioned in 10 sub-clusters, using a semi-automated method summarized in Appendix 2, in order to group specific categories of symptoms, like \textit{polyuria, drowsiness and tirediness, eye problems} etc. A further partition  has been performed on the subtopic \textit{negative feelings}, in order to distinguish between \textit{anger, sadness, stress} and \textit{fear}.

These topics and sub-topics are at the basis of the data analytic phase presented  in Section \ref{sec:phenotype}.

\subsection{Patients segmentation by gender and age}
\label{subsec:genderage}
In all examined health blogs, patients are anonymous and have no IDs. Therefore, it is impossible even to understand if a patient has repeatedly participated in a discussion. However, for about a 25-30\% of the threads, questions to doctors and discussions on symptoms and perceptions are reported by a patient's relative or friend, rather than by the patient itself. This can be captured by  key phrases, like "my mother", "my husband", and the like. This information is precious, since it allows, at least for a fragment of messages in our collection, to perform patients' segmentation according to age and gender. For example, "my mother" can with good confidence be attributed to an elder female diabetic patient. 

First, to filter from the collection only messages authored by some relative of a diabetic patient,  we selected  discussion threads that mentioned a parental relation (e.g., "my father", "my sister", etc.)  within the first fifty words. In fact, most of the time, we observed that when a user writes about his/her relative, the relationship is clarified in the first lines of a message or thread.  
   
The evaluation of this classification process was conducted in a way similar to that described in Section \ref{subsec:filtering}, obtaining an accuracy of 0.93. Details are omitted for the sake of space. 

\begin{figure}[ht]
  \centering
  \includegraphics[width=1.1\linewidth]{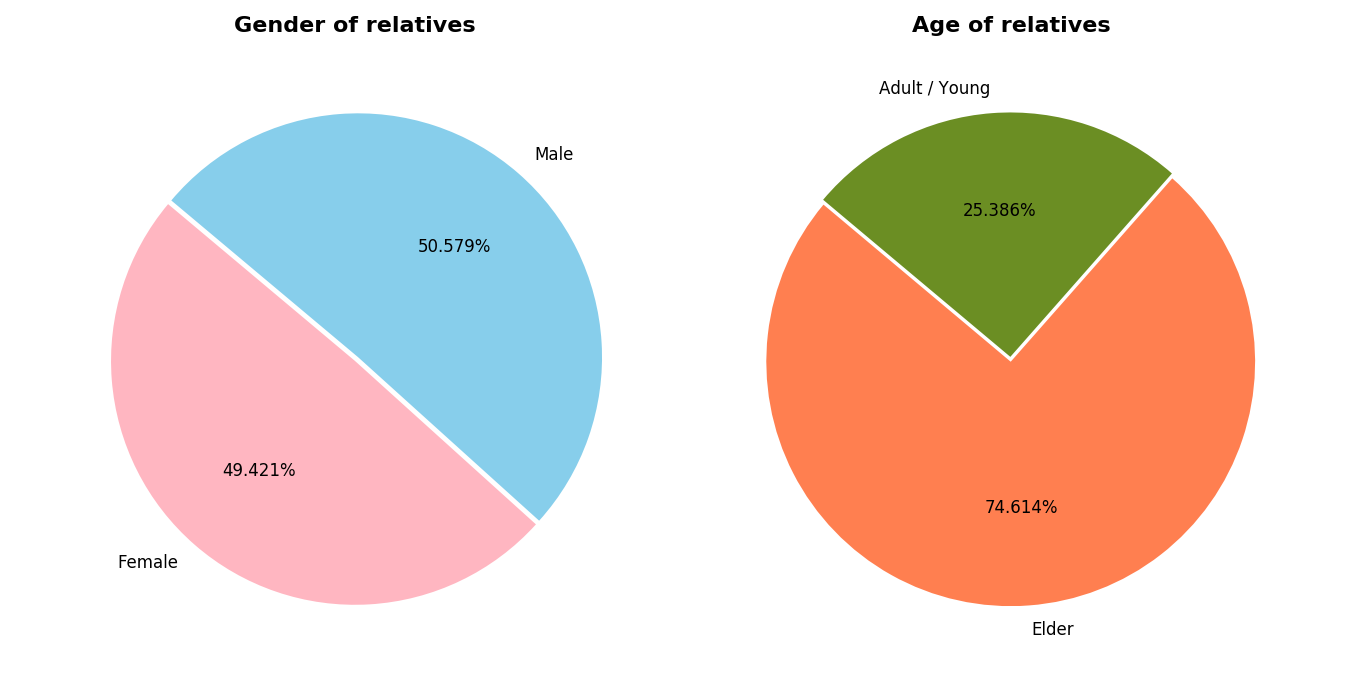}
  \caption{\textit{\small Gender and age of patients who are relatives of a thread's author}}
  \label{fig:genderAge}
\end{figure}

For patient segmentation on this subset of data, 
we associated parental keywords in messages to two age groups, as follows:
\begin{itemize}
\item \textbf{old}: grandfather, grandmother, father, mother, uncle, aunt, father-in-law, mother-in-law;
\item \textbf{adult/young}: husband, wife, boyfriend, girlfriend, brother, sister, cousin, brother-in-law, sister-in-law, son-in-law, daughter-in-law,son, daughter, nephew.
\end{itemize}
Gender can be easily inferred for the majority of these parental keywords.
Figure \ref{fig:genderAge}  shows some general statistics on gender and age. It shows that, in this sample of our collection, patients are almost equally distributed according to gender, while  most of the times (74.51\%) the patient whose symptoms are reported by a parent author, are elder, as expected. 

\section{Data Analytics:  The social phenotype of diabetes}
\label{sec:phenotype}

Data analytics in healthcare is evolving into a promising field that takes advantage of the increasing daily availability of health data to extract insights for producing better-informed decisions \cite{ref:24}. 
In this Section our objective is to show that, by analyzing the social phenotype of patients, \textit{it is possible to derive insightful information on symptoms and complications that, in spite of their clinical severity, concern patients the most, and negatively affect their lives and their well-being}. This information is relevant also because it is provided by patients in a spontaneous way, according to their priorities, rather than  in response to specific questions formulated by specialists (as for in Quality of Life assessments).

 In what follows, data are analyzed in the light of  research questions presented in Section  \ref{intro}.

\subsection{R1: What are the main complications reported by patients for selected diseases?  Which symptoms  are more frequently reported by patients?} 
\label{complications}
Figure \ref{fig:frequencyOfTopics} shows the frequency of the 14 main discussion topics listed in Appendix 1. 
As already anticipated in Section \ref{subsec:topic}, 9 of these topics were found to correspond to common diabetes complications and risk factors \cite{complications}\cite{complications2},  precisely: \textit{glycemia, hepathology/gastroenterology, gynecology, angiology, cardiovascular complications, neurology, ophtalmology/orthopedics, andrology} and \textit{endocrinology}. 
The other topics are: \textit{symptoms/feelings, dietary issues, clinical tests and drugs, diabetes in general, various problems}. 

 In \cite{complications} diabetic peripheral neuropathy (corresponding to our topic  \textit{neurology}) is indicated as the most common complication, affecting from 30 to 50\% of diabetic individuals. The other most common complications are associated with cardiovascular problems (topics \textit{cardiology} and \textit{angiology}),  kidney (topic \textit{hepathology}), foot and eye (topic \textit{opthalmology/ortopedics}). This order of relevance, however,  is not reflected in Figure \ref{fig:frequencyOfTopics}. The histogram shows that patients are eager to discuss their personal feelings, worries, and prevailing symptoms (topic \textit{symptoms and feelings}) in the vast majority of messages. Besides symptoms, the most frequent topics concern  glycemic complications (note in Appendix 1 that this topic also includes insuline and insuline-related terms), dietary problems, clinical tests and drug effects.  Although these problems are not considered by diabetologists as representing a severe limitation of a patient's life\footnote{except for the assumption of insulin, which is known to be a major cause of discomfort},  at least in comparison with other symptoms affecting eyes, limbs and kidneys \cite{complications2},  Figure \ref{fig:frequencyOfTopics}  shows that indeed they are reported by patients as being a major cause of  discomfort.

\begin{figure}
  \centering
  \includegraphics[width=1\linewidth]{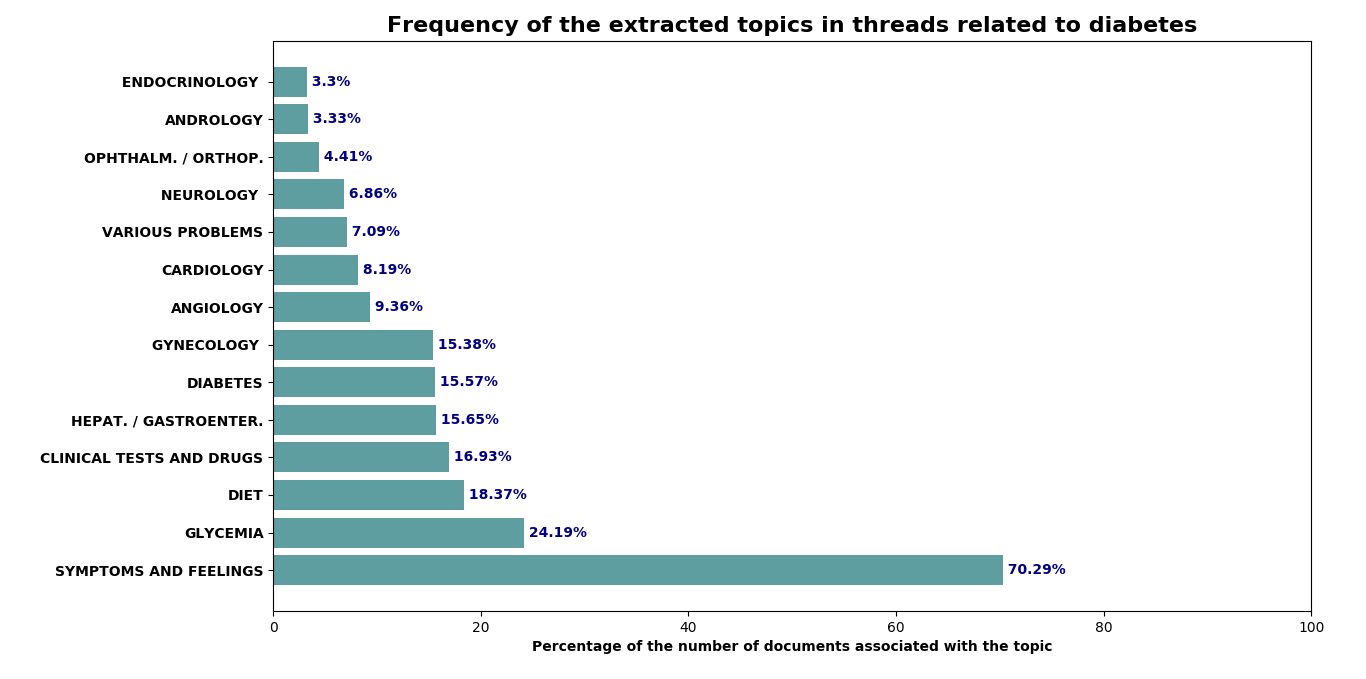}
  \caption{\textit{\small Relevance of extracted topics in threads concerning diabetes}}
  \label{fig:frequencyOfTopics}
\end{figure}

\begin{figure}
  \centering
  \includegraphics[width=1.05\linewidth]{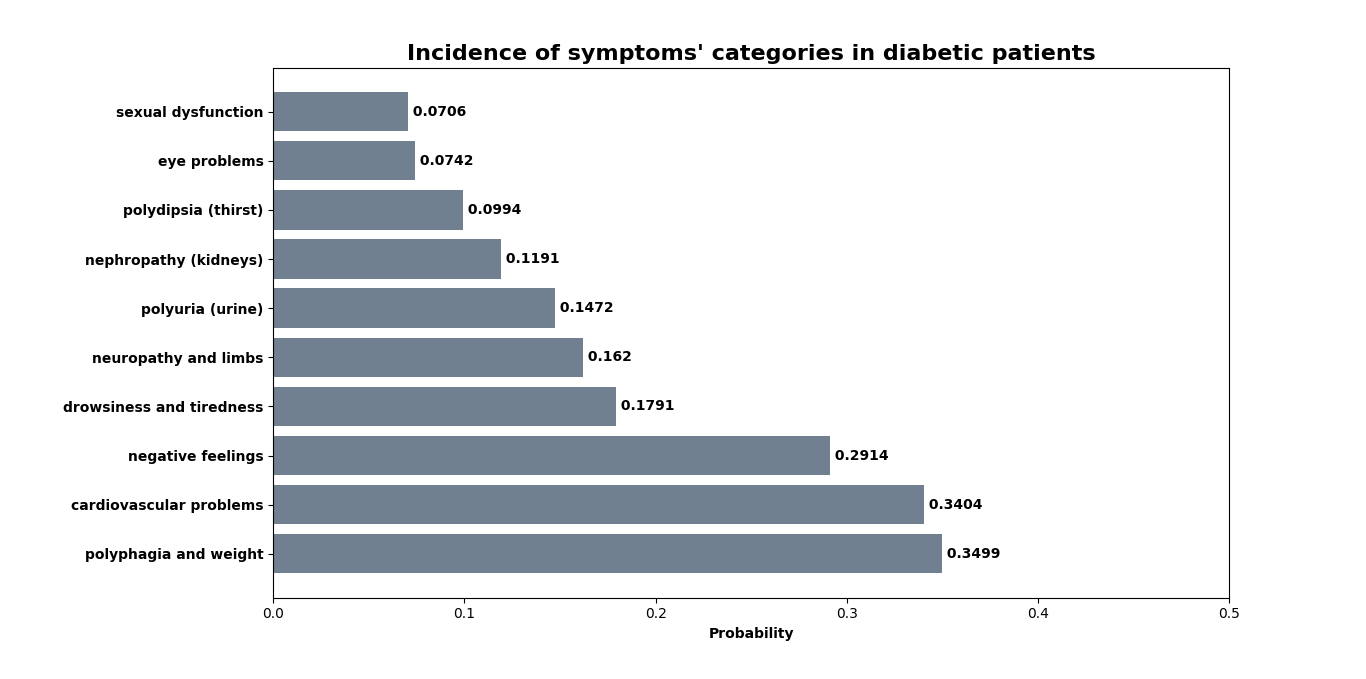}
  \caption{\textit{\small Percentage of occurrence of "symptoms and feelings" sub-topics in all threads }}
  \label{fig:symptoms}
\end{figure}

\begin{figure}
  \centering
  \includegraphics[width=1.05\linewidth]{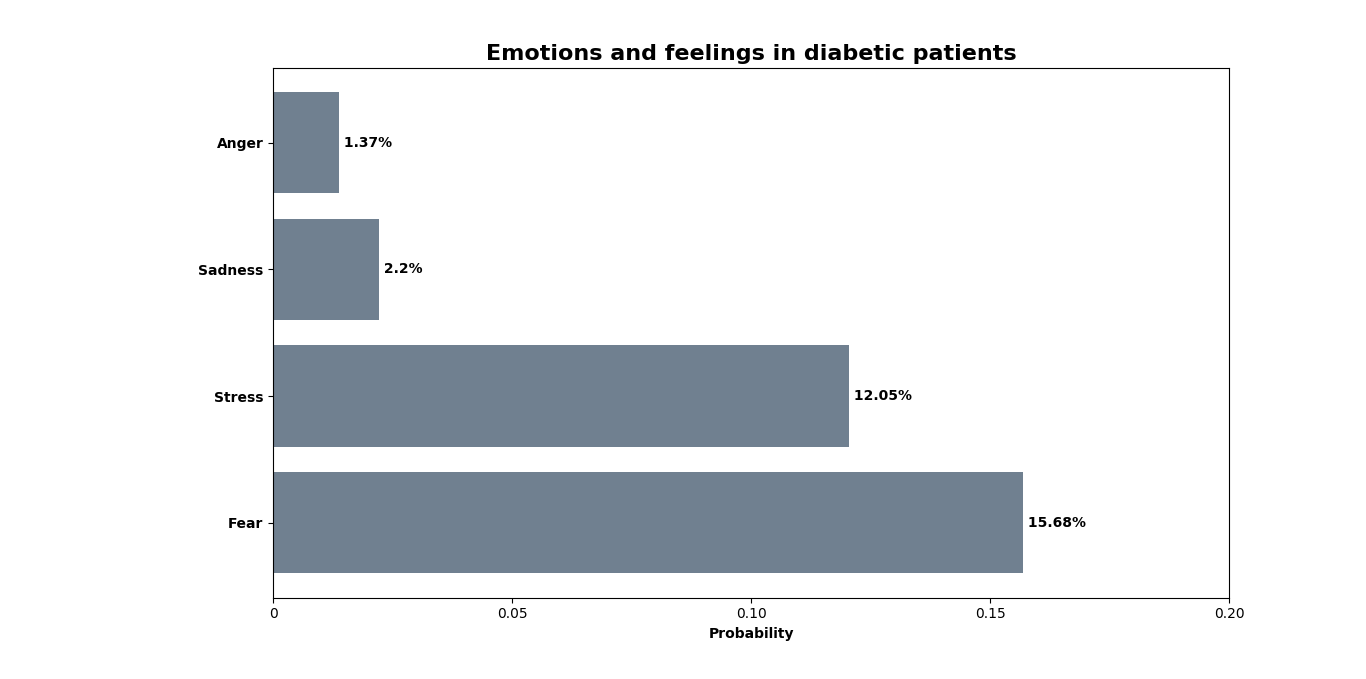}
  \caption{\textit{\small Relative incidence of "negative feelings" types  }}
  \label{fig:emotions_and_feelings}
\end{figure}

Since \textit{symptoms and feelings} is the most frequent and populated topic, and it is also the most interesting in order to outline the social phenotype of diabetes,  we further partitioned it into 10 subtopics, using a semi-automated procedure described more in detail in Appendix 2. The 10 sub-topics are: \textit{negative feelings, polyuria, polydipsia, polyphagia and obesity, drowsiness and tirediness, eye symptoms, cardiovascular symptoms, nerve and limbs symptoms, sexual disfunctions, nephropaty}. Negative feelings have been further partitioned into: \textit{anger, sadness, stress, fear}.  
The histograms in Figures \ref{fig:symptoms}  and \ref{fig:emotions_and_feelings}   show the distribution of these sub-topics.

\begin{figure}
  \centering
  \includegraphics[width=\linewidth]{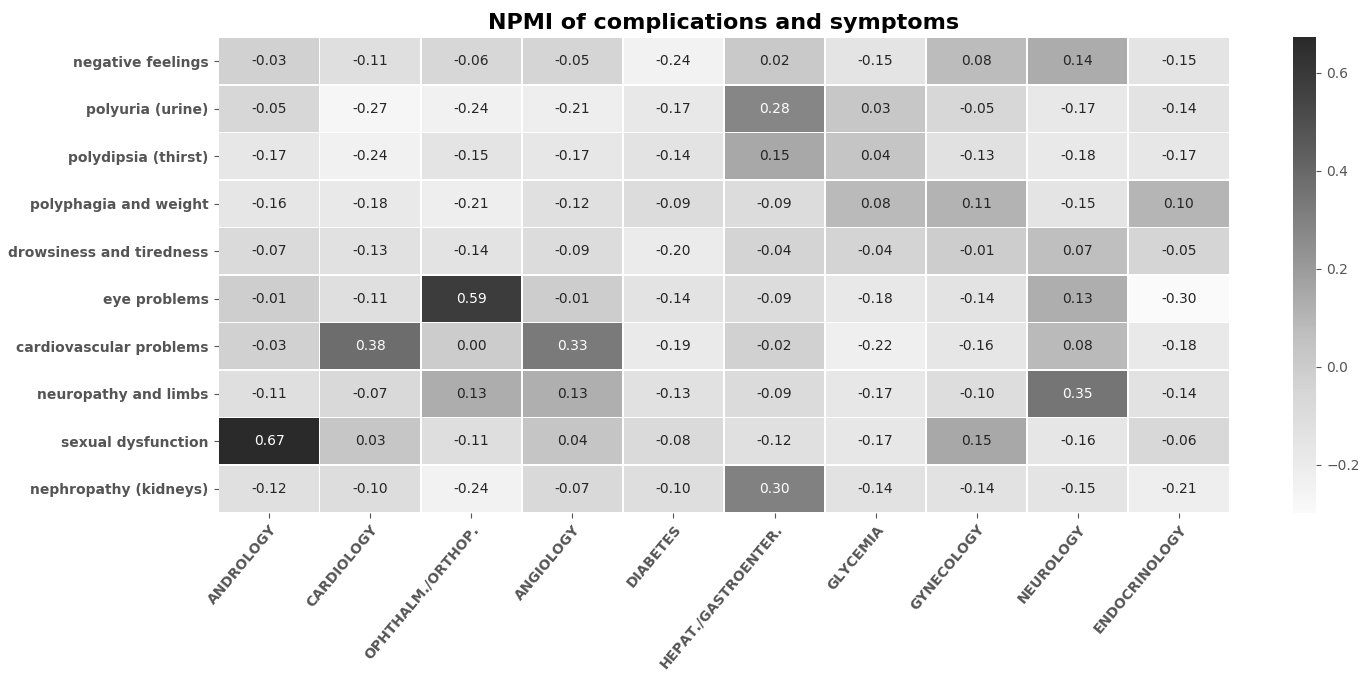}
  \caption{\textit{\small Correlation between symptoms/feelings types and complications (Normalized Point-wise Mutual Information) 
  }}
  \label{fig:feelings-complications}
\end{figure}

According to the latent topic modeling approach adopted in this paper, every message is a mixture of topics. We are here interested in understanding the correlation strength between these topics, and in particular, \textit{what types of symptoms are complained in relation with what types of complications}. 
This is shown in the heatmap of Figure \ref{fig:feelings-complications}. To study the correlation, we use the \textit{Normalized Pointwise Mutual Information, NPMI}\footnote{\url{https://pdfs.semanticscholar.org/1521/8d9c029cbb903ae7c729b2c644c24994c201.pdf}}.
In general, symptom types show the higherst NPMI values for related  complications, as expected. For example,   “eye problems” such as: \textit{blurred, out of focus, shadow, tarnished..} are the most frequently complained symptoms in patients discussing ophthalmology/ orthopedics  complications (\textit{retinopathy, diabetic retinopathy, maculosis, etc.}). Eating disorders or concerns (polyphagia and weight) are mentioned especially in relation with glycemic, gynecological and endocrinological complications. 
\textit{Negative feelings} are pervasive in all messages, therefore they do not show a strong correlation with any specific complications, although neurology is shown to  have some degree of correlation.

\subsection{R2: Are there gender/age differences in a disease perception?} 
\label{segmentation}
\begin{figure}
  \centering
  \includegraphics[width=1\linewidth]{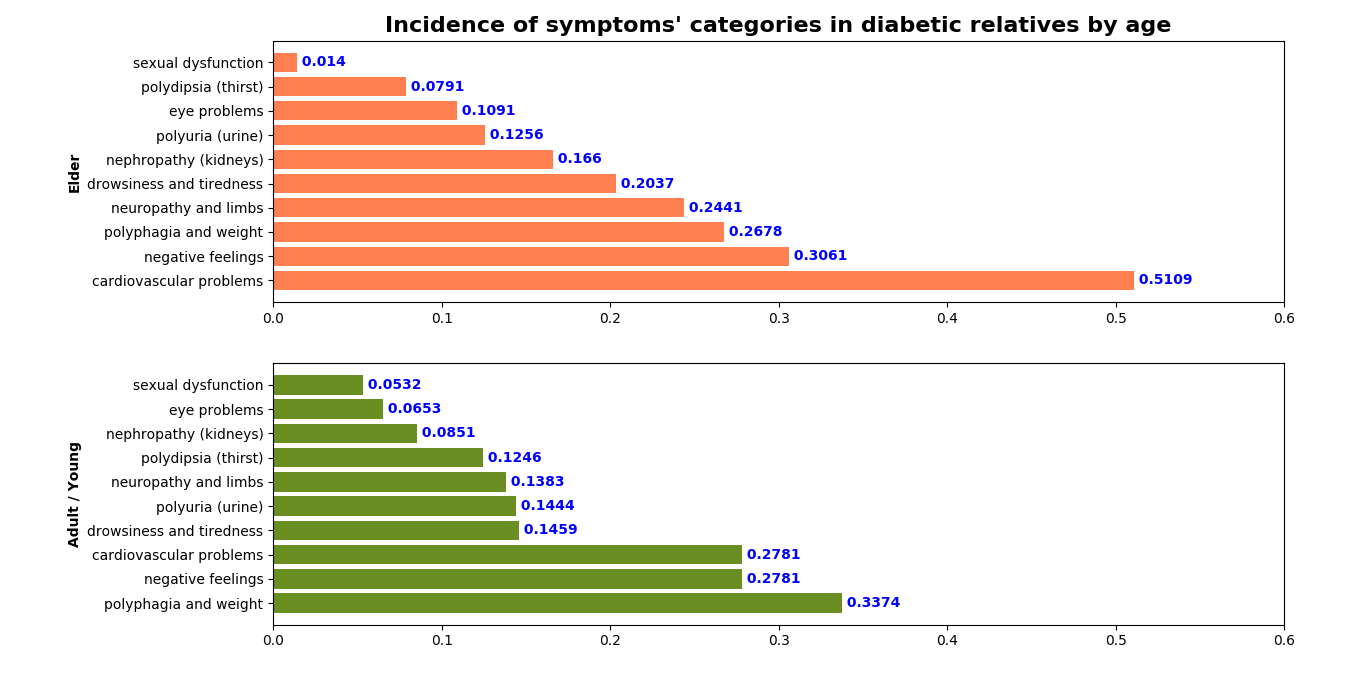}
  \caption{\textit{\small Percentage of occurrence of "symptoms and feelings" sub-topics in age-profiled messages}}
  \label{fig:symptoms-age}
\end{figure}

To answer this question, we considered our age and gender profiled sample of the data. In analyzing this data, we must keep in mind that, although presumably the  patients' symptoms and complications are objectively reported by the authors of messages (i.e., \textit{the patients' relatives}), it is not possible to attribute \textit{negative feelings}   to  either the patients or their relatives. \\
 When considering the distribution of  symptoms’ categories by age (Figure \ref{fig:symptoms-age}),  we note that elder patients are more frequently reported to suffer or fear cardiovascular problems (tachycardia, heart attack, stroke) while adults are more concerned with eating disorders and overweight.  Furthermore, when comparing Figure \ref{fig:symptoms-age} with the distribution of symptoms in the total population (Figure \ref{fig:symptoms}), we may notice that the latter is more similar to  the Adult/Young histogram than to the Elder histogram. This makes sense, since we can imagine that adults and young, more than elder patients, may decide to express their concerns on health-related on-line forums. Elders "speak" in the words of their younger relatives.

\begin{figure}
  \centering
  \includegraphics[width=1\linewidth]{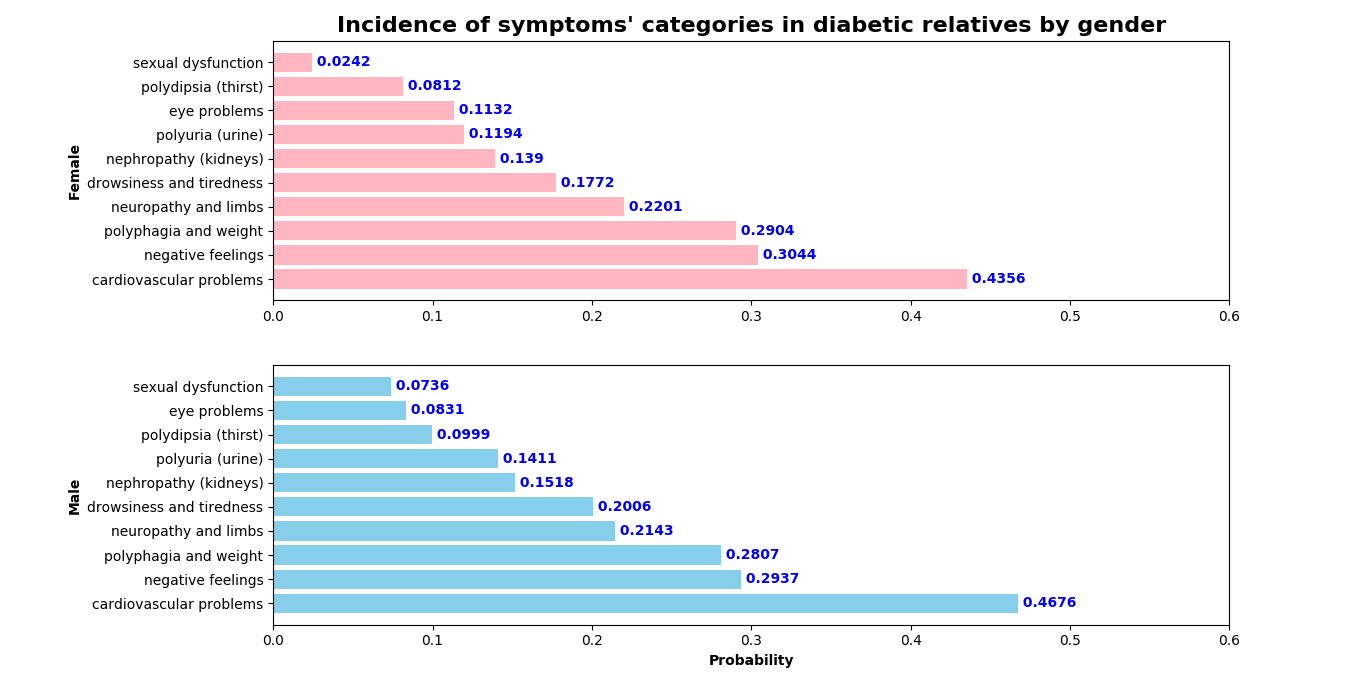}
  \caption{\textit{\small Percentage of occurrence of "symptoms and feelings" sub-topics in gender-profiled messages }}
  \label{fig:symptoms-gender}
\end{figure}

 When analyzing gender, we note  no significant difference between males and females (Figure \ref{fig:symptoms-gender}), except than for sexual disfunctions, which are mentioned  significantly more frequently in association with male patients than females, although the incidence of these symptoms is low in both gendered samples.
However, if we consider previous Figure \ref{fig:feelings-complications} on symptoms by complications, a gender difference can be  noted when comparing messages discussing about andrological and gynecological problems. Patients with andrological disorders (presumably males) strongly fear about their sexuality, while patients with gynecological disorders (presumably females) show correlation with  polyphagia and weight, sexual disfunctions (much less than males) and negative feelings.  


\subsection{R3: Does this "patient-centric"   study of a disease perception correspond to published results using traditional Quality of Life assessments?} 
\label{surveys}

\begin{figure}
  \centering
  \includegraphics[width=1\linewidth]{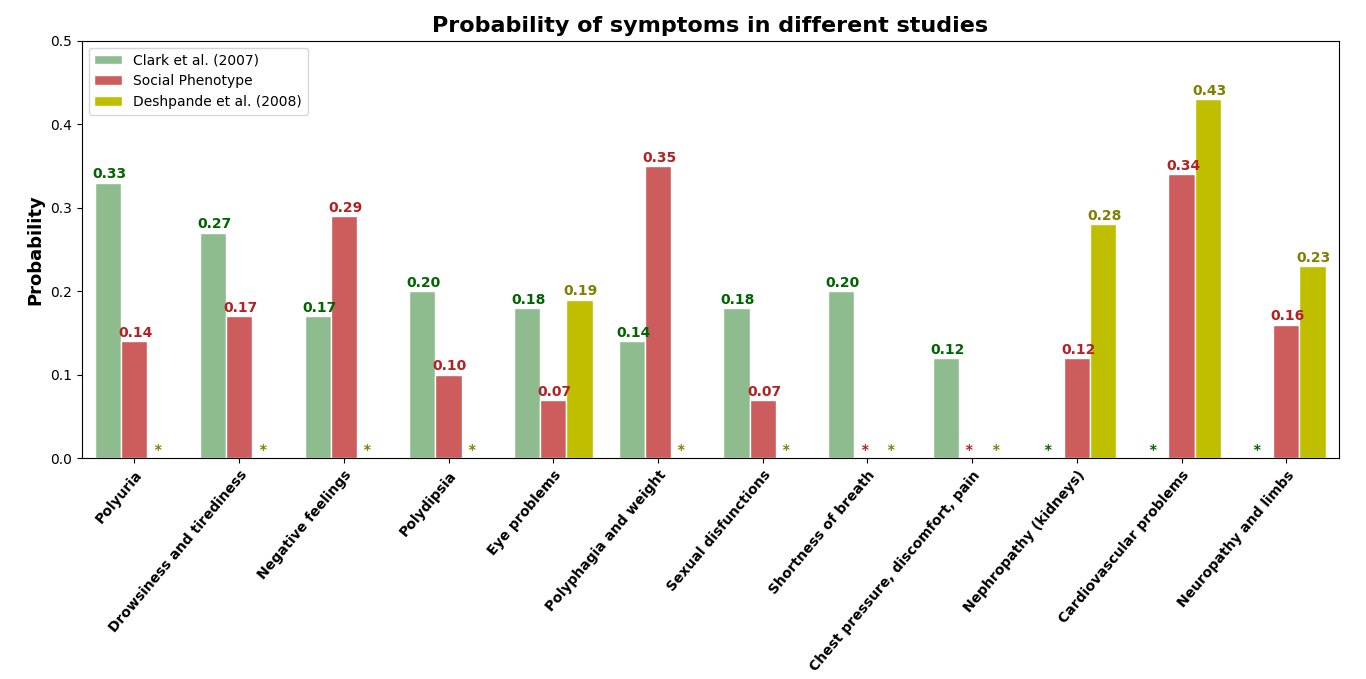}
  \caption{\textit{\small Comparison of social phenotype of diabetes with two survey studies  \cite{complications}\cite{complications2}}}
  \label{fig:comparison}
\end{figure}
 
In the past years there has been considerable research on health-related quality of life (QoL) assessments, motivated by the fact that clinical variables alone do not fully capture patients' perception of their health \cite{diabetesqualityoflife}. Many of these studies focused on diabetes, like \cite{diabetesqualityoflife2017} \cite{complications}\cite{complications2} and several others, due to the high diffusion of this disease.  The typical way in which these studies are conducted is by administering a questionnaire to a sample of diabetic patients with different features, e.g., with and without diabetic complications, type 1 o type 2 diabetes, demographics, etc. However, due to the difficulty of retrieving a  large sample of patients,
these studies have been able to consider only a subset of possibly influencing variables, in order to obtain sufficient statistical evidence. 

A popular questionnaire is the MDQoL-17 \cite{MDQoL} \cite{diabetesqualityoflife2017}  \cite{Faye}. Questions typically include  frequency of pain types, worry types (e.g., missing work or diabetes complications), and satisfaction with social relationship, sexual life, current treatment, etc. The results however are rather contradictory. For example, the authors in \cite{diabetesqualityoflife2017} say that patients with Type 2 diabetes have a negative impact on their quality of life with or without complications. Instead,  \cite{Faye} found that the presence of complications or comorbidity conditions in Type 2 patients
has  the most significant influence on
QoL of diabetic patients. In this regard, \cite{MDQoL} observe that: "because the selection of
items was driven by the statistical
item relationships provided by the
participants in this study, the extent
to which these participants and their
responses are not representative of
the general population of people with
diabetes may limit the generalizability
of the instrument".  In other terms, the small number of responders is the main limitation of the majority of QoL studies. 

A much larger-scale experiment was presented in \cite{complications2}. They present  a 5-year observational study of
individuals with or at risk for diabetes diagnosis, in which  15,794 questionnaires were returned. In this case, evidence was sufficient to partition responders in 4 different categories (low risk, high risk, type 1, type 2) and 6 demographic categories. The baseline survey consisted
of 64 questions on health-related
topics, including symptoms, comorbidities,
family history, etc.   A total of 13 symptom types were included in the questionnaire. The authors however report that, although some of these symptoms should be frequent is any diabetic patient, they were \textit{not} reported by about 70\% of the respondents:
"it will be important to determine whether this large pool of diagnosed but largely asymptomatic respondent [..] reflect earlier diagnosis and adequate symptom control through appropriate treatment \textbf{or simply attention to the wrong symptoms}" \cite{complications2}.

This is exactly the main added value of experiments like the one reported in this paper: \textit{patients are freely expressing their symptoms and concerns}, without any bias possibly injected by doctors. Furthermore, symptoms' types are automatically learned from patients' generated texts, using topic detection techniques, rather than being defined \textit{a priori} by doctors - a practice that does not facilitate a comparison among different studies, since different questionnaires may list  different symptoms\footnote{compare, e.g., \cite{complications2}  and \cite{complications}}. 
The second advantage is the dimension of the sample, that can easily reach very high numbers. The obvious disadvantage is that patients cannot be profiled in any way. Although, in future extensions of the present work, more refined text mining techniques can be used for larger-scale profiling, detailed profiling such as in \cite{complications2}  cannot be achieved. 

Due to the previously outlined diversity of methodological instruments, a comparison among the results of this study and state-of-the-art QoL assessment is not fully appropriate, but we nevertheless attempted to compare our analysis on most frequently reported symptoms (Figure \ref{fig:symptoms}) with analogous statistics reported in  \cite{complications2}  and \cite{complications}, two studies in which the main considered symptoms are explicitly listed.  To facilitate a comparison among different symptom labels, 
we manually identified associations (when appropriate) between  symptoms' labels used in these surveys and the ones used in our study. For example, we merged statistics on symptoms:  "extreme hunger" and "unusual weight loss" in \cite{complications2} and paired these symptoms with our sub-topic "polyphagia and weight". Similarly, we aggregated \textit{heart attack, chest pain, coronary heat disease, congestive heart failure,  stroke} in \cite{complications} and paired with our topic "cardiovascular problems".  When reported data in surveys where partitioned according to patients' demographic of disease categories (e.g., type 1 or type 2 diabetes), we computed a weighted average of the reported values.   The result of this comparison is shown in Figure \ref{fig:comparison}.  Although statistics are rather different, it is worth noticing that symptoms types observed in our study correspond to those considered in survey studies, and that our study is more complete in this sense, because our list is the union of all symptoms reported in either  \cite{complications} or \cite{complications2}\footnote{note that shortness of breath and chest pressure are listed in \cite{complications2}, but although related words have been found in our topics, there was no easy way to associate these two symptom labels with our topic labels.}. We further note that dietary and weight issues are those for which there is the larger difference between survey studies and our social phenotype. It may seem surprising, but   probably  food restrictions constitute a daily and constant limitation of diabetic patients' everyday  life, and though not in itself serious, it is experienced with much discomfort.

\section{Concluding remarks}
\label{sec:conclusion}
In this paper we presented a study on diabetic patients' quality of life, based on their freely expressed descriptions of the disease, symptoms and complications. 
Our study was conducted on a large sample of threads extracted from 4 popular Italian health blogs. Text classification and a state-of-the-art topic detection techniques have been used to derive what we called the "social phenotype" of a disease. Although the idea of extracting the digital phenotype of diseases (a broader concept that includes the social phenotype) has been launched in \cite{ref:3}, to the best of our knowledge, this is the first large-scale study in which the utility of this approach is demonstrated.

The main findings of our study can be summarized as follows:
\begin{enumerate}
    \item The  most frequent topics   discussed by patients are their negative feelings (which are pervasive in almost all messages), followed by (hyper-hypo)glycemia and insulin-related issues,  dietary issues, and clinical tests. Often, these issues - especially diet restrictions and clinical tests - are not considered in QoL questionnaires, since they are not perceived by doctors as representing severe limitations. Our data show that instead, diet,  glycemic control, drugs and clinical tests have a negative and daily impact on patients' life, and they are discussed more frequently than common and  severe diabetic complications \cite{complications2}, such as problems with  kidneys, eyes and limbs. 
    \item Elder patients mostly fear, or suffer, cardiovascular symptoms (frequent keywords are: tachicardia, beating..)  while adult and younger patients are concerned with polyphagia and alimentary disorders (frequent keywords are: hungry,  appetite, put on  weight,  lose  weigh...).
    \item We found no significant difference between females and males,  however females affected by gynecological complications fear overweight and express discomfort (negative feelings), while males affected by andrological complications are mostly worried about sexual disfunctions.   
\end{enumerate}

Because it is impossible to accurately profile patients, the social phenotype of a disease is complementary, not substitutive, of traditional QoL questionnaires. However, it presents a number of advantages with respect to state-of-the-art survey studies:
\begin{enumerate}
    \item The dimension of the sample is much larger in comparison with responders of questionnaires. In the present study we paid much attention to the precision of classified messages (see Section \ref{subsec:filtering}) in spite of recall, and in addition, we restricted to health blogs in Italian, which eventually led us to collect a dataset of 40443 patients' messages. Extending this study to English blogs could easily increase the dataset by one order of magnitude, or more. Note however that the Italian sample (as remarked in Section \ref{sec:data}) represents the typical profile of industrialized nations.
    \item Survey studies can hardly be compared with each other, because the types of considered symptoms in questionnaires may be different, and demographic profiles of patients are also different. Instead, the social phenotype "objectively" collects and cluster complications, worries, symptoms and feelings, just as they are expressed in the patients' words. It reflects the reality of patients' everyday life, and how they perceive their disease, without any preconception or influence injected by medical knowledge. This has led to a number of surprising findings, but also to a number of expected findings (for example, 9 of the automatically detected topics were found to  correspond to main known diabetes complications), to confirm the accuracy of the adopted text mining techniques.
\end{enumerate}

\section*{Acknowledgments}
\DipInfHack

\bibliographystyle{mybibfile}

\bibliography{mybibfile}

\appendix
\section*{Appendix 1: Lists of topics}
\label{appendix1}
The 14 extracted topics are listed below in descending order from the most common to the least predominant in the collection. 
For all the topics, only the most significant terms are reported in order of importance. All terms are translated from Italian to English for readability. Furthermore,  since words are represented in LDA as multinomial probability distributions over topics, \textit{there are overlapping words in different topics}.

Topic \textit{labels} have been manually assigned, however  no manual correction of terms in topics has been performed: although topics may include some apparently less related terms, they  reflect similarity of contexts according to the  topic model described in Section \ref{subsec:topic}. For example, note that  terms such as \textit{insulin, infection}  and \textit{surgery}, although they are neither symptoms nor feelings, have been associated to the topic SYMPTOMS-FEELINGS.  
Since topics are generated by contextual similarity and co-occurrence, this clearly shows that patients are emotionally affected by, and fear,  insulin, infection  and surgery. 

\begin{itemize}

\item \textbf{(ID=1) \textit{SYMPTOMS-FEELINGS}}: [\textit{symptoms, pain,  weight, pressure, insulin, anxiety, stress, fear, diet, glycemia,  obese, low, leg, burning, annoyance, feeling, fatigue, feeding, attack, tiredness, nausea, overweight, cramp, malaise, dizziness, fat, weariness, tachycardia, crisis, hypertension, sleep, slimming, sweat, hyperglycemia, swelling, sting, metabolism, heart, psychology, cardiac, metformin, surgery, hypoglycemia, hair, erection, cold, eyes, tremor, infection, skin, inflammation, effort, thyroid, panic, nervous, hyper-cholesterol}];

\item \textbf{(ID=2) \textit{GYNECOLOGY}}: [\textit{gynecology, cycle, ovary, pregnancy, pill, hormonal, polycystic, polycystic ovary, gynecology obstetrics, prolactin, ovule, progesterone, fsh, abortion, testosterone, menstruation, tsh, dosage, follicle, acne, estradiol, hormone, amenorrhea, endocrinology, insulin resistant, micropolicistic, insulin, hormone dosage, beta, micropolicistic ovary, pregnant, hirsutism, follicle, metformin, cortisol, acid folic, ovary, mui, contraceptive, insulinemic, menopause, thyroid, dhea, ovary syndrome, delay, pcos, endometrium}];

\item \textbf{(ID=3) \textit{DIET}}: [\textit{diet, food, meal, fruit, vegetable, carbohydrate, sugar, bread, ferritin, lunch, meat, fat, hunger, digestive endoscopy, gastroenterology endoscopy, sweet, breakfast, milk, biscuit, nutrition, fish, integral, lymphocytes, globules, neutrophils, coffee, protein, dinner, consuming, diarrhea, monocyte, eosinophile, iron, foods, oil, weight, slice, anemia, intolerance, celiac disease, glycemia, mch, mcv, rice, cheese, acid, vitamin, mchc, fruit vegetable, honey, platelets}];

\item \textbf{(ID=4) \textit{GLYCEMIA}}: [\textit{glycemia, curve, glucose, diabetology, fasting, metabolism, diabetology metabolism, glycemic curve, basal, insulin, fasting blood sugar, hemoglobin, load curve, diabetes, glycated, insulinemic, glucose metabolism, glycated hemoglobin, gestational, gestational diabetes, diabetes metabolism, glucose charge, analysis, basal glycemia, diet, glycosylated, meal, glycosylated hemoglobin, sweet, insulin glycemia, weight, glycaemia analysis, fasting glucose, urine, diabetes blood sugar, diabetes endocrinology, blood sugar, sugar, high, blood sugar insulin, low blood sugar, nutrition, glycemic load, diet weight, basal insulin, breakfast, blood test, glucose urine, hyperglycemia}];

\item \textbf{(ID=5) \textit{CLINICAL TESTS}}: [\textit{cholesterol, triglycerides, hdl, analysis, cholesterol hdl, ldl, blood, creatinine, urine, transaminase, gamma, gpt, bilirubin, high, high cholesterol, proteins, got, cholesterol triglycerides, azotemia, hemoglobin, glycemia, tot, blood analysis, ast, ldl cholesterol, cholesterol glycemia, hdl triglycerides, leukocytes, psa, glucose, albuminine, hematology, blood cholesterol, ketones, uric, hdl ldl, nephrology, uric acid, uricaemia, statin, globulin, ves, amylase, blood count, cholesterol analysis, got gpt, urinalysis, potassium, urea, alpha, creatininemia, ggt, glycated, urobilinogenesis, phosphatase}];

\item \textbf{(ID=6) \textit{HEPATOLOGY/GASTROENTEROLOGY}}: [\textit{hepatology, gastroenterology, liver, hepatic, urine, liver hepatology, gastroenterology liver, renal, pancreas, pancreatic, abdomen, ultrasound, kidney, steatosis, calculus, urinary, urination, tumor, cirrhosis, tac, bladder, gallbladder, nephrology, prostate, kidney, lesion, liver cirrhosis, focal, liquid, abdominal, spleen, lung, abdomen ultrasound, diameter, fatty liver disease, infiltration, insipid diabetes, cysts, colic, bile, biliary, infection, lymph node, carcinoma}];

\item \textbf{(ID=7) \textit{ANGIOLOGY}}: [\textit{cardiology, artery, stenosis, carotid, hypertension, ecg, angiology, aortic, valve, vascular, coronary, heart, mitral, ventricular, vascular surgery, plaque, atheromasia, carotid, calcification, ischemia, echo-doppler, cardiac surgery, doppler, heart attack, dilated, cardioaspirin, diabetes hypertension, vascular angiology, echo-cardiogram, electrocardiogram, cardiac, flow, coronarography, thickening, angioplasty, venous, vessels, parietal, femoral, anterior, hypertensive, cardiopathy, chest, mellitus, systole, diabetes mellitus, bilateral, smoker, arterial hypertension, hypertrophy, occlusion, vertebral, cardiovascular}];

\item \textbf{(ID=8) \textit{DIABETES}}: [\textit{diabetes, diabetes endocrinology, diabetes endocrinology, mellitus, metformin, diabetology, diabetes mellitus, lunch, diabetic, dinner, insulin, glycemia, dinner lunch, microalbuminuria, curve metabolism, diabetes blood sugar, lantus, diabetes diet, breakfast lunch, breakfast, metforal, prandiale, diet, glycemic level, diabetes insulin, hypoglycemia, glucophag, meal, glycated, basal insulin, fasting, mini-curve, pellets, peptide, nutrition, weight, diabetes analysis, fasting glycemia, sugars, obesity, insulin-dependent, glycated hemoglobin, weight height, cpr, metabolic, proteinuria, postprandial, machine, dose, weight loss, pressure, hemoglobin, hyperglycemia, novonorm, kcal, glucose, carbohydrate, cholesterol, insulin resistant, triglycerides, glycosyl, triatec, glicid, insipid, simvastatin, ramipril, lasix}];

\item \textbf{(ID=9) \textit{VARIOUS\_PROBLEMS}}: [\textit{dermatology, disability, cortisone, venereology, dermatology venereology, legal, psychiatry, cataract, anesthesia, drops, cream, radiotherapy, hernia, mellitus, transplant, diabetes mellitus, plant, aneurysm, fever, depression, neurosurgery, smoking, bacterium, reanimate, cutaneous, lumbar, carrier, immunology, xanax, apnea, hypoglycemic, remedy, allergology, recognize, coma, aggravated, allergy, diabetes, psoriasis, antihistamine, marrow, oral, abilify, polyneuropathy, transfusion, sleepy, lesion, neuropathy, orthopedics, heart disease, carcinoma, immunosuppressant, Alzheimer}];

\item \textbf{(ID=10) \textit{NEUROLOGY}}: [\textit{neurology, neuropathy, pain, tingling, resonance, encephalon, orthopedics, ischemia, rmn, cerebral, geriatrics, ankle, arthritis, tac, magnetic, rheumatology, leg, fingers, magnetic resonance, nerve, arm, limb, knee, rheumatoid, diabetes, dementia, Parkinson, diplopia, stroke, vascular, ventricular, hand, ambulate, atrophy, sub-cortical, back, arthrosis, occlusion, unstable, dizziness, frontal, hypertension, bilateral, head turning, memory, brain, swelling, hip, injury, cephalus, physiotherapy, vascular, motor, suffering, amplitude, hypertensive, fracture, muscle, limb, articular, micro, heart disease, skull, ear, anterior, microcyst, femur, ophthalmology, cranial, emisom, paresthesia, mesencephalon, occipital, cerebellum, hemisphere}];

\item \textbf{(ID=11) \textit{CARDIOLOGY}}: [\textit{heart, cardiovascular, cardiovascular heart, pressure, heart attack, aorta, respiratory, heart breathing, hypertensive, cardioaspirin, cardiologist, heart disease, coronary, diabetic hypertension, hypertension, diabetes, triatec, erectile dysfunction, norvasc, glycosuria, angioplasty, artery, cardiac, vasculopathy, angina, plavix, arterial pressure, coronography, stent, senile, ictus, torvast, low pressure, prevention, diabetes, rhythm, anticoagulant, pharmacological, hereditary, bypass, ischemia, cerebral, lobivon, mellitus, diabetes mellitus, prevention, cardiovascular, peripheral, arterial hypertension, smoking, cholesterol, insulin diabetes, atheromasia, coronary artery disease, hypertension, hypercholesterolemia, arteriopathy, dyslipidemia, ima, arteriosclerosis, dilatrend, coumadin, ticlopidin, lopresor, ramipril, enapren, congescor, tenormin, aprovel, cardirene, atenolol, nebilox, blopress, ascriptin, micardis, antipertens, cerebrovascular, cardura, seloken}];

\item \textbf{(ID=12) \textit{OPHTHALMOLOGY/ORTHOPEDICS}}: [\textit{eyes, oculist, oculus, foot, retina, retinopathy, diabetic retinopathy, laser, maculosis, wound, injection, degeneration, myopia, surgery, edema, hallux, circumcision, glaucoma, fluorangiography, plastic, amputation, insulin mellitus, thrombosis, optical, cicatrization, dark, fistulas, procedure, opening, patch, puncture, surgery, fingers, local, lenses, calf, radius, see, operative, vitreous, close, blind, canal, opaque, wrist, resorption, healing, dot, swelling, vitreous, temporary, ischemia, myodesopsie, holes, hole, iris, cornea, perforator, corneal, scar, cataract, vitrectomy, crystalline, suture, fissure, flap, gangrene, peripheral}];

\item \textbf{(ID=13) \textit{ENDOCRINOLOGY}}: [\textit{nutrition, biochemistry, hormone glands, gland, hormone, endocrinology, endocrinology glands, slimming, sideraemia, thyroid, hypothyroid, eutirox, dietology, plasmaicot, basal blood sugar, weight gain, glucose glucose, tsh, mcg, hyperthyroid, glycemic load, insulin glycemia, thyroid endocrinology, diet, psychotropic, kcal, weight, phosphorus, chemiluminescent, diet, tapazol, absorption, diet nutrition, glycemic curve, insulinemic curve, cortisol, intake, weight height, hct, chili, diabetes endocrinology, tpo, glycemia analysis, integral, insulin, prandial, resistant insulin, resistant, thyroglobulin, glycemia diet, pharmacology, hunger, bilirubin, tirosint, tiroxin, levotiroxin, metformin, triglycerides, cholesterol, ths, ferritin, parathormone, tyrosine, prolactin, hypocaloric, prolactinemia, cholesterolemia, metforal, dosage, metabolism, glycoto, homocysteine}];

\item \textbf{(ID=14) \textit{ANDROLOGY}}: [\textit{andrology, erection, sexual, glans, cialis, testosterone, ejaculate, viagra, impotent, urology, libido, morning, masturbation, critical, hyperglycemia, decline, sex, occasional, inhibitors, male, partner, erectile, reach, ace, emotional, timetable, precocious, nocturnal, psychology, rigid, approach, depression, extreme, auto-erotic, excited, orgasm, erectile, vardenafil, misfire, sildenafil, tadalafil, virile, enduring, eroticism, lust, embrace, vigor, excitatory}].

\end{itemize}
\section*{Appendix 2: Sub-topics of the topic "Symptoms"}
\label{appendix2}

The topic SYMPTOMS-FEELINGS is the most important for the purpose of characterizing the social phenotype of diabetes. For this reason, words in this cluster have been further partitioned into 10 sub-topics and enriched in a semi-automated manner:   

\begin{enumerate}
\item First, we manually identified some semantically related terms of the initial SYMPTOMS-FEELINGS topic,  creating ten small initial clusters (each cluster consisted of four or five terms).
\item With Word2Vec, we obtained the vector spaces representing the starting clusters.
\item With Word2Vec, we selected the terms of the collection most similar to the starting clusters (with a similarity score of at least 0.80 w.r.t. the cluster vector spaces) and we added these words to the ten clusters.
\end{enumerate}
The ten symptom-feelings sub-clusters are\footnote{as before, labels have been added manually, and cluster members have been translated in English}:

\begin{itemize}
\item \textbf{\textit{NEGATIVE FEELINGS}}: [\textit{fear, anxiety, stress, worry, pain, frustration, depression, unnerved, terror, dissatisfaction, melancholy, discomfort, paranoia, anger, sadness, loneliness, anguish, bad mood, impatience, restlessness, insecurity, unhappy, disappointment, discouragement, drama, exasperated, overwhelmed, worn down, unmotivated, resentment, hatred, sad, apprehension, repressed, displeasure, tragedy, malaise, angry, restless, oppressive, destabilized, crying}];
\item \textbf{\textit{POLYURIA (URINE)}}: [\textit{urine, polyuria, pis, pee, peeing, urination, mingle, bladder, pollakiuria, nocturia, voiding, dripping, dysuria, hematuria}];
\item \textbf{\textit{POLYDIPSIA (THIRST)}}: [\textit{polydipsia, thirst, drought, drink, glass, liter, sip, rasp, water, cup, drink, beverage, swallow, gulp}];
\item \textbf{\textit{POLYPHAGIA AND WEIGHT}}: [\textit{polyphagia, hungry, appetite, put on weight, lose weight, overweight, underweight, obese, weight, fat, thin, pounds, skinny, kilo, heavy, slender, waist, belly, hyper-caloric, greedy, body, skinny, kcal, hypo-caloric, glutton, BMI, long-limbed, decay, eat}];
\item \textbf{\textit{DROWSINESS AND TIREDNESS}}: [\textit{sleep, sleepy, tired, weak, exhausted, dazed, heavy, dizzy, very weak, breathless, fatiguing restlessness, listlessness, weakness, perennial, malaise, nausea, numbness, heartbeat, weakened, listless, out of breath, breathless, tiring, weary, jaded, worn}];
\item \textbf{\textit{EYE PROBLEMS}}: [\textit{eye, retina, retinopathy, tarnished, blurred, sight, low vision, visually impaired, blind, monocular, maculosis, maculopathy, visus, sight, glaucoma, scotoma, monocle, shadow, glow, out of focus, amblyopia, cataract, vitreous}];
\item \textbf{\textit{CARDIOVASCULAR PROBLEMS}}: [\textit{beating, tachycardia, hypertension, pressure, heart attack, stroke, heart, cardivascular, thrombosis, artery, vein, cardiopathy, thrombus}];
\item \textbf{\textit{NEUROPATHY AND LIMBS}}: [\textit{neuropathy, nerves, peripheral, limb, foot, leg, hand, amputation, forearm, calf, thumb, wrist, little finger, arm, elbow, ankle, big toe, thigh, paraesthesia, heel, palm, finger}];
\item \textbf{\textit{SEXUAL DYSFUNCTION}}: [\textit{dysfunction, erectile, erection, sexual, impotent, viagra, cialis, levitra, libido, libidine, sildenafil, tadalafil, vardenafil, orgasm, eroticism}];
\item \textbf{\textit{NEPHROPATHY (KIDNEYS)}}: [\textit{nephropathy, kidney, renal, microalbuminuria, proteinuria, creatinine, hydronephrosis, glomerulus, albuminuria, protenuria, creatininemia, nephritis, monorenes, GFR, microhematuria, chronic renal failure, bacterium, glomerulonephritis, nephrology, clearanec, pielectasia, pielonefrite}].
\end{itemize}

\end{document}